\documentclass[aps,prl,twocolumn,amssymb,showpacs]{revtex4}
\usepackage[dvips]{graphicx}

\begin{document}

\title{Nonequilibrium dynamical ferromagnetism of interacting Single-Molecule Magnets}

\author{Gui-Bin Liu and Bang-Gui Liu}
\email[Corresponding author:~]{bgliu@aphy.iphy.ac.cn}
\affiliation{Institute of Physics, Chinese Academy of Sciences, Beijing 100190, China\\
Beijing National Laboratory for Condensed Matter Physics, Beijing 100190, China}

\date{\today}

\begin{abstract}
We propose a nonequilibrium Monte Carlo (MC) approach to explore
nonequilibrium dynamical ferromagnetism of interacting single
molecule magnets (SMMs). Both quantum spin tunneling and thermally
activated spin reversal are successfully implemented in the same
MC simulation framework. Applied to a typical example, this
simulation method satisfactorily reproduces experimental
magnetization curves with experimental parameters. Our results
show that both quantum and classical effects are essential to
determine the hysteresis behaviors. This method is effective and
reliable to gain deep insights into SMMs.
\end{abstract}


\maketitle

{\it Introduction.} The single-molecule magnet (SMM) is very
interesting because of its potential applications in spintronic
devices\cite{spintr1,spintr2}, quantum computing\cite{qcomp1},
information storage\cite{mem}, optical transistors\cite{optran},
and so on. The most famous examples are
Mn$_{12}$\cite{Mn12a,Mn12b}, Fe$_8$\cite{Fe8}, and
Mn$_4$\cite{Mn4a}. A SMM usually has well-defined spin ground
state and large magnetic anisotropy, and can exhibit hysteresis
loops at low enough temperatures\cite{smmbook}.
Landau-Zener\cite{Landau,Zener} (LZ) model has been used to
explain the steps of the hysteresis loops\cite{Mn4b,Fe8b,smmbook}.
On theoretical side, numerically exact method has been used to the
phenomenological spin model for the Fe$_8$ and Mn$_{12}$ systems,
and calculated results show that LZ model is reliable for usual
experimental field sweeping rates in describing the two-level
problems at the level crossing fields\cite{numeric1,numeric2}, but
experimental hysteresis loops cannot be explained because
thermally activated processes are not considered\cite{smmbook}.
Thermally activated processes, however, play important roles in
determining the nonequilibrium magnetization dynamics and even
level populations. A quantitative method which is able to
reproduce the full hysteretic magnetization dynamics of SMMs is in
need.

Here, we propose a nonequilibrium Monte Carlo (NQMC) approach
taking both the LZ quantum tunneling effect and the thermally
activated processes into account, and hereby satisfactorily
reproduce and explain experimental magnetization behaviors of
SMMs.

{\it Spin model.} For simplicity without losing physics, we
arrange SMMs on a finite two-dimensional (2D) rectangular lattice
of $L_1\times L_2$, and consider inter-SMM magnetic dipolar and
exchange interactions. Actually, some single layers of patterned
SMMs has already been grown on specific solid
surfaces\cite{onSi,onAu,onsur}, and it has been proved that a SMM,
when put on a Au surface, can still keep its essential
properties\cite{mem}. We use giant spin approximation. Every SMM
has a spin of $S$. We use a typical lattice constant $a=1.5$\,nm,
which is an intermediate value of well-known SMMs. The model
Hamiltonian reads
\begin{equation}
H=\sum_i H_i^0 + \frac 12\sum_{i\ne j}{\vphantom\sum} (J_{ij}^{\rm
di}-J_{ij}) \mathbf{S}_i\cdot\mathbf{S}_j \label{eq.H}~,
\end{equation}
where $J_{ij}^{\rm di}$ and $J_{ij}$ describe the magnetic dipolar
and exchange interactions between spins at $i$ and $j$. The factor
$1/2$ is due to the double counting in the summation. The first
term describes the part for all the single SMMs, and $H_i^0$ is
given by
\begin{equation}
H_i^0=-D_2(\hat{S}_i^z)^2-D_4(\hat{S}_i^z)^4+H_i^{\rm tr} + g\mu_B
B_z\hat{S}_i^z~, \label{eq.Hi0}
\end{equation}
where $D_2$ and $D_4$ are positive anisotropic parameters, $g$ the
Land\'e g-factor (here $g\!=\!2$ is used), $\mu_B$ the Bohr
magneton, $B_z(t)$ the external magnetic field in the $z$
direction. $\hat{S}_i=(\hat{S}_i^x,\hat{S}_i^y,\hat{S}_i^z)$ is
the quantum spin operator for the $i$th SMM. As for the transverse
term $H_i^{\rm tr}$, it is usually defined as $H_i^{\rm tr}=E[
(\hat{S}_i^x)^2 - (\hat{S}_i^y)^2 ] + g\mu_B B_x\hat{S}_i^x$,
where $E$ is the second-order transverse anisotropic constant and
$B_x$ is the transverse external field in the $x$ direction. As
usual, $B_x$ is a constant and $B_z(t)$ is a linear function of
the time $t$ with the sweeping rate $\nu$.

{\it LZ tunneling probability.} The easy axis determines two
equilibrium spin directions in the $\pm z$ direction. When
calculating the LZ tunneling probability, we treat the SMM spins
coupled with the $i$th spin by using the mean field approximation.
As a result, for the $i$th SMM, we have the effective one-SMM
Hamiltonian
\begin{equation}
H_i=H_i^0+g\mu_B B_i^{\rm eff} \hat{S}_i^z
\label{eq.Hi}
\end{equation}
where the effective mean field $B_i^{\rm eff}$ is given by
\begin{equation}
B_i^{\rm eff}=\sum_{j(\ne i)}(J_{ij}^{\rm di}-J_{ij})S_j^{\rm
eq}/(g\mu_B)~, \label{eq.Bieff}
\end{equation}
where $S_j^{\rm eq}$ is the equilibrium value ($S$ or $-S$) for
the $i$th spin along the easy axis. This approximation is natural
in the NQMC simulation, as will be clarified in the following.
Then the standard diagonalization technique can be used to solve
Eq.~(\ref{eq.Hi}). Hamiltonian Eq.~(\ref{eq.Hi}) has $2S+1$ energy
levels, which can be labeled by the quantum numbers
$m=S,S\!-\!1,\cdots,-\!(S\!-\!1),-S$ to a first-order
approximation. If without the transverse part $H_i^{\rm tr}$,
Eq.~(\ref{eq.Hi}) is diagonal, and there are level crossings at
some special field values. When the transverse term $H_i^{\rm tr}$
is taken into account, the level crossings become
avoided\cite{Mn4b,Fe8b,smmbook}. When $B_z$ is swept close to
$B_{m,m'}$ at which the avoided level crossing happens between
states $m$ and $m'$, quantum tunneling occurs between the two
states. Strictly speaking, this tunneling is beyond LZ model which
is a two-states theory\cite{Landau,Zener}, but LZ tunneling can be
used as a good approximation for each of the tunneling processes
because LZ transition time is usually very short compared to time
increments between the two successive avoided level crossings
\cite{numeric1,numeric2,LZtime}. The nonadiabatic LZ tunneling
probability $P_{m,m'}$ is given by\cite{Landau,Zener}
\begin{equation}
P_{m,m'}=1-\exp\Big[-\frac{\pi\Delta_{m,m'}^2}{2\hbar g \mu_B
|m-m'|\nu}\Big]~, \label{eq.pLZ}
\end{equation}
where the tunnel splitting $\Delta_{m,m'}$ describes the energy
gap at the avoided crossing of states $m$ and $m'$. $B_{m,m'}$ and
$\Delta_{m,m'}$ can be calculated by diagonalizing
Eq.~(\ref{eq.Hi}).

{\it Thermally activated reversal rate.} In order to calculate
thermal spin reversal rate, we use classical spin approximation
which is reasonable and reliable because the spin usually is very
large for SMMs. As a result, the energy of the $i$th SMM is given
by
\begin{equation}
E_i=-D_2(S_i^z)^2-D_4(S_i^z)^4+h_iS_i^z \label{eq.Hip}
\end{equation}
where $h_i=g\mu_B(B_z+B_i^{\rm eff})$. Each of the spins has two
equilibrium orientations ($\pm S$) along the easy axis. We use the
angle $\theta_i$ to describe the $i$th spin's deviation from its
original equilibrium orientation. All the other angle values
($0<\theta_i<\pi$) are the transition states for the spin to
reverse its orientation. We express $S_i^z$ as $S_i^{\rm
eq}\cos\theta_i$. Usually, there exists a maximum in the curve of
$E_i(\cos\theta_i)$, and the maximum determines the energy barrier
for the spin reversal \cite{liying,lbg}. Defining
$x_i=\cos\theta_i$, we have $-1\leq x_i \leq 1$. Then the barrier
is given by
\begin{equation}
\Delta E_i=
\left\{\begin{array}{lcl}
E_i(x_i=x_i^m) &,&{~\rm if~} |x_i^m|\leq 1\\
E_i(x_i=-1)=2|h_iS_i^{\rm eq}| &,&{~\rm if~} x_i^m<-1~,\\
E_i(x_i=1)=0  &,&{~\rm if~} x_i^m>1
\end{array}\right.
\label{eq.dEi}
\end{equation}
where $x_i^m$ (according to the maximum) is defined as
\begin{equation}
x_i^m=\sqrt[^{^{^3}}\!\!]{-\frac{q}2+\sqrt{\Delta_d}} + \sqrt[^{^{^3}}\!\!]{-\frac{q}2-\sqrt{\Delta_d}}~,
\label{eq.xim}
\end{equation}
where $\Delta_d=(q/2)^2+(p/3)^3$, $p=D_2/(2D_4S^2)$, and
$q=-h_iS_i^{\rm eq}/(4D_4S^4)$. Then the reversal rate of the
$i$th spin is determined by the Arrhenius law\cite{ArrheniusLaw1},
$R_i=R_0\exp({-\Delta E_i/(k_B T)})$, where $k_B$ is the Boltzmann
constant and $R_0$ the characteristic attempt frequency.

\begin{figure}[t]
\begin{center}
\includegraphics[width=8cm]{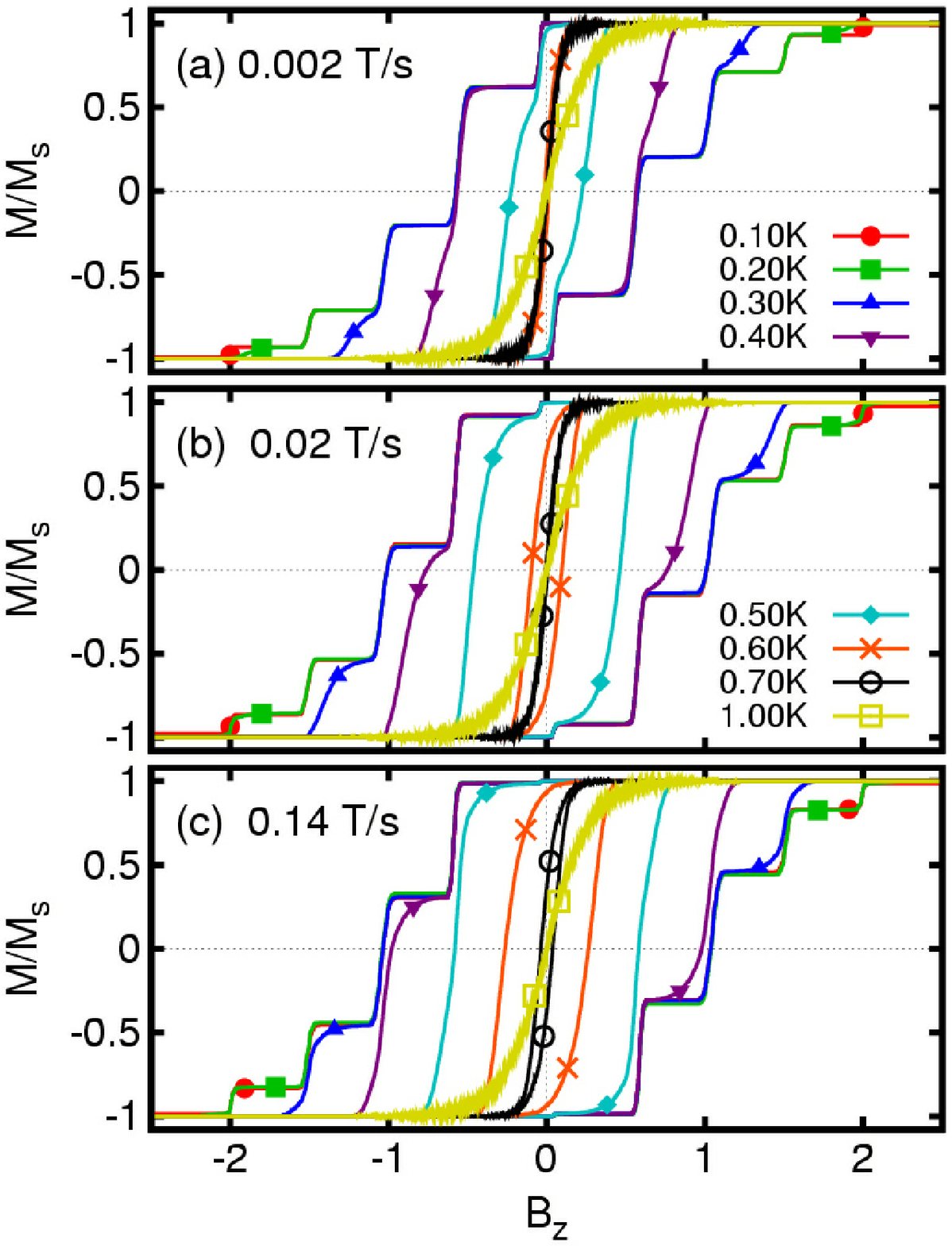}
\caption{(color online). Simulated magnetization curves ($M/M_s$
vs. $B_z$ in T) with three sweeping rates: $\nu=0.002$ (a), 0.02
(b), and 0.14 (c)\,T/s. For all the three panels, the temperatures
are $T=0.10$ ({\Large $\bullet$}), 0.20 ({\small $\blacksquare$}),
0.30 ($\blacktriangle$), 0.40 ($\blacktriangledown$), 0.50
($\blacklozenge$), 0.60 ({\large $\times$}), 0.70 ({\Large
$\circ$}), and 1.00 ({\small $\square$})\,K.} \label{fig.diffT}
\end{center}
\end{figure}

{\it Simulation method and parameters.} To set up our NQMC
simulation, we define MC steps by the time points, $t_n=\Delta t
\cdot n$, where $n$ nonnegative integers. Every spin takes either
$S$ or $-S$ at each of the times $t_n$. A spin can be reversed by
overcoming the thermal barrier and tunneling through LZ mechanism.
The probability $P_i$ for a thermally activated spin reversal
within a MC step is defined by $P_i=\Delta t\cdot
R_i$\cite{liying,lbg}, and the LZ tunneling probability is given
by Eq. (\ref{eq.pLZ}). Actually, we have the magnetization $m=S$
at the beginning of field sweeping, and then need to consider only
the avoided crossings at $(S,m')$, where $m'=-S,-S+1,\cdots$. When
the field is swept to $B_{S,m'}$, we consider the LZ tunneling by
adding the LZ probability $P_{S,m'}$ to the thermally activated
reversal probability. Because our Monte Carlo time step is much
larger than the LZ transition and spin relaxation times, a spin
will have long enough time to transit to the possible lowest level
at current field after a LZ tunneling is finished\cite{superrad}.
Therefore, our NQMC simulation is self-consistent.

For convenience in comparing with experiment, we choose the Mn$_4$
system to demonstrate the NQMC method and its power. Hence we have
$S=9/2$, $D_2/k_B=0.608$K, $D_4/k_B=3.8$mK, and $E/k_B=32$mK from
Ref. \onlinecite{Mn4b}. We assume that our 2D lattice is put on a
metal surface and hence use a carrier-mediated 2D interaction for
$J_{ij}$ ($J/k_B=5.8$mK for the nearest spins)\cite{2drkky}. This
inter-SMM interaction, weak enough, does not destroy the essential
SMM properties. In addition, we already show that simulated
results are not sensitive to specific choices of the interaction,
and other interactions yield similar results. We take $B_x=0.02$T,
$\Delta t=1$ms, and $R_0=10^9$/s. Such parameter choices guarantee
the good balance between computational demand and precision. In
our simulations, field $B_z$ is swept from -2.5T to 2.5T in the
forward process, and the full hysteresis loop is obtained simply
by using the loop symmetry. Every hysteresis loop is calculated by
averaging over 100 runs to reduce statistical errors.

{\it Simulated results and discussions.} We have done our NQMC
simulations with various ($L_1$,$L_2$) values and experimentally
accessible sweeping rates $\nu$ (0.002$\sim$0.14 T/s) and
temperatures $T$ (0.10$\sim$1.00 K)\cite{Mn4b}. Presented in Fig.
1 are our simulated magnetization $M$ (normalized to the saturated
value $M_s$) curves against the applied sweeping field $B_z$ with
$L_1\!=\!L_2\!=\!10$ for three sweeping rates, namely 0.002, 0.02,
and 0.14 T/s. For each of the three cases, we present the results
for eight temperatures: 0.10, 0.20, 0.30, 0.40, 0.50, 0.60, 0.70,
and 1.00 K. It is clear that there exist hysteresis loops for low
enough temperatures and complete paramagnetic magnetizations are
obtained for high enough temperatures. We can clearly see four
steps between 0 and 2T for low enough temperatures. In addition,
we have done NQMC simulations for various values of $L_1$ and
$L_2$. For square lattices ($L=L_1=L_2$), the step heights
decrease a little with increasing $L$ from 8 to 50. For
rectangular lattices with $L_1\times L_2=24^2$, the step heights
change a little for different $L_1$ and $L_2$. It is clear that
the step structures are far from those predicted by LZ tunneling
alone.

In fact, thermally activated reversals play important roles in
determining the magnetization curves. Our simulations show that
the number of the steps decreases with increasing the temperature
for given value of the sweeping rate $\nu$. It can be seen that
the area enclosed by a hysteresis loop increases monotonically
with increasing the sweeping rate and decreasing the temperature,
and the smallest visible loop is at 0.70 K for $\nu=0.14$ T/s,
0.60 K for 0.02 T/s, and 0.50 K for 0.002 T/s, respectively. The
thermal effect increases substantially when we increase the
temperature from 0.3K on. There will be no step structures for
usual sweeping rates when the temperature is higher than 0.5K. The
hysteresis loop will disappear when the temperature is too high.

{\it Conclusion.} We have proposed a NQMC approach to explore the
nonequilibrium dynamical ferromagnetism of interacting SMMs. Both
quantum spin tunneling and thermally activated spin reversal are
successfully implemented in the same Monte Carlo simulation
framework. Applied to the Mn$_4$ system, this simulation method
satisfactorily reproduces experimental magnetization curves with
experimental parameters. Our results show that both quantum and
classical effects are essential to determine the hysteresis loops
in the magnetization curves. This method is effective and
reliable, and can be used to gain deep insights into SMMs and
other quantum nanomagnets with strong anisotropy.

This work is supported  by Nature Science Foundation of China
(Grant Nos. 10874232 and 10774180), Chinese Department of Science
and Technology (Grant No. 2005CB623602), and Chinese Academy of
Sciences (Grant No. KJCX2.YW.W09-5).

\end{document}